\documentclass[preprint,12pt,chicago,numbers,sort&compress]{elsarticle}

\usepackage{graphicx} % Required for inserting images
\usepackage{amssymb}
\usepackage{hyperref}
\usepackage{appendix}
\usepackage{amsmath}
\usepackage{mathrsfs}
\usepackage{caption}
\usepackage{subcaption}
\usepackage{geometry} % 加载geometry宏包
\usepackage{braket}
\usepackage{color}
\usepackage{natbib}

\hypersetup{colorlinks = true, %将链接文字带颜色
	    linkcolor = blue, %图表引用颜色设置为蓝色
	    urlcolor = blue,%网页链接为蓝色
            citecolor = blue} %文献引用颜色设置为蓝色

\journal{Physics Letters A}
\newenvironment{acknowledgment}
{\section*{Acknowledgment}}
{}
\newenvironment{Declaration of competing interest}
{\section*{Declaration of competing interest}}
{}

\begin{document}

\begin{frontmatter}


\author[1]{Ruoyang Cui}
\ead{2201310220@stu.xjtu.edu.cn}

\author[2]{Yaojin Li \corref{cor1}} 
\ead{20230029@lut.edu.cn}

\address[1]{School of Physics, Xi'an Jiaotong University, Xi'an 710049, China}
\address[2]{Department of Physics, School of Science, Lanzhou University of Technology, Lanzhou 730050, China}

\cortext[cor1]{Corresponding author}


\title{Longitudinal optical conductivity of graphene in van der Waals heterostructures composed of graphene and transition metal dichalcogenides}


%\author{}

%\affiliation{organization={},%Department and Organization
  %          postcode={}, 
  %          state={},
   %         country={}}

\begin{abstract}
Placing and twisting graphene on transition metal dichalcogenides (TMDC) forms a van der Waals (vdW) heterostructure. The occurrence of Zeeman splitting and Rashba spin-orbit coupling (SOC) changes graphene's linear dispersion and conductivity. Hence, this paper studies the dependence of graphene's longitudinal optical conductivity on Rashba SOC, the twist-angle and temperature. At zero temperature, a main conductivity peak exists. When Rashba SOC increases, a second peak occurs, with both extremes presenting an enhanced height and width, and the frequencies where the two peaks arise will increase because the energy gap and the possibility of electron transition increase. Altering the twist-angle from 0 to $30^{\circ}$, the conductivity is primarily affected by chalcogen atoms. Moreover, when temperature increases to room temperature, besides a Drude peak due to the thermal excitation, a new band arises in the conductivity owing to the joint effect of the thermal transition and the photon transition.
\end{abstract}




\end{frontmatter}


\section{Introduction}
\label{sec:intro}
%介绍材料（背景知识）
%石墨烯很牛逼，转角石墨烯很牛逼
Van der Waals (vdW) heterostructures are a heterojunction formed by stacking two-dimensional (2D) materials with weak interlayer interactions\cite{geim2013van,novoselov20162d,jariwala2017mixed}. In recent years, an important research direction is the twisted van der Waals 
heterostructure, with the graphene/transition metal dichalcogenide (TMDC) heterostructure being one of the most common ones \cite{wang2015electronic,di2017angle,hou2017robust,lu2017moire,du2017modulating,cao2018unconventional,cao2018correlated,rahman2019utilization}. 
As reported in 2004\cite{novoselov2004electric}, graphene is a pure carbon 2D material with a hexagonal lattice. Its most stimulating property is the Dirac point near the Fermi level, which has a linear dispersion and can be described using the Dirac equation\cite{neto2009electronic}, allowing for a wide range of applications in electronics and optoelectronics\cite{stankovich2006graphene,dikin2007preparation,kim2009large,mueller2010graphene}. One of its extraordinary characteristics is the high optical conductivity\cite{falkovsky2007space,falkovsky2007optical,peres2007phenomenological,hanson2008dyadic,falkovsky2008optical}. On both the experimental and theoretical fronts, transport
measurements have found excellent agreement with the low-energy effective model\cite{neto2009electronic,beenakker2008colloquium,sarma2011electronic}. Meanwhile, TMDCs have a wide bandgap\cite{fang2015ab} and thermal stability\cite{zhang2015measurement,zhang2017phonon,yu2020plane}, thus contributing in important in energy conversion applications\cite{li2015emerging,li2018metallic} and other fields\cite{mcdonnell2016atomically,anju2021biomedical}. This graphene/TMDC heterostructure also has special advantages, such as electrical and optical responses\cite{zhao2018electrically,rahman2019utilization}.

%石墨烯并不是那么牛逼
The existence of Rashba spin-obit coupling(SOC) and Zeeman splitting in graphene has changed the linear property\cite{kane2005quantum,min2006intrinsic,yao2007spin} and led to a wide range of new physical phenomena\cite{haldane1988model,kane2005quantum}. Therefore, it is of vital necessity to investigate the impact of Rashba SOC and Zeeman splitting. 
%Except for the characters above, graphene has also contributed to the advancement of two-dimensional materials, including transition metal dichalcogenides(TMDCs)\cite{mas20112d,gupta2015recent,akinwande2017review,ares2022recent}. 
%When graphene is stacked on another type of 2D material, a heterostructure forms\cite{novoselov20162d,jariwala2017mixed}.
It has been reported that one can continuously modulate the above mentioned quantities by rotating the heterostructure\cite{li2019twist,david2019induced}, which provides a new method to change graphene's energy bands and conductivity. It should be noted that the Diarc point still exists in the energy gap of TMDCs\cite{ma2011first,kaloni2014quantum,gmitra2015graphene,agnoli2018unraveling}.
% 别人做过的工作（很相关）

The dependence of the optical conductivity on different physical observables, such as the chemical potential, the light polarization, uniaxial strain and correlations
%and bilayer graphene\cite{tabert2012dynamical,liu2015nonlinear,xie2023phase} 
has been studied thoroughly\cite{scholz2013interplay,li2013vanishing,ang2014nonlinear,liu2014topologically,sadeghi2015anisotropic,zubair2020influence,alidoust2022controllable,xie2022far,xie2023phase,huang2023nonlocal}.
%The influence on the optical conductivity from spin splitting caused by the magnetization has been studied thoroughly\cite{alidoust2022controllable}. 
%When calculating the influence from the chemical potential, the similarity of the Hamiltonian between bilayer graphene with an asymmetry gap and single layer graphene with Rashba SOC and Zeeman splitting can be used\cite{nicol2008optical}. 
%还差的东西
For example, in ref. \cite{li2013vanishing}, the low frequency peaks in the optical conductivity is compared to the peaks in the joint density of states when taking both Rashba and Dresselhaus SOC into account. Ref. \cite{xie2022far} has shown the effect of uniaxial strain on the optical conductivity of graphene and ref. \cite{xie2023phase} has illustrated Rashba SOC's impact on gated bilayer graphene's conductivity.
However, the dependence of the optical conductivity of the graphene/TMDC heterostructure on the values of Rashba SOC and temperature has not been investigated. Besides, the effect on graphene's optical conductivity from the twist-angle in the graphene/TMDC heterostructure is still unclear.

%文章结构
Therefore, we investigate in detail the impact of the physical observables on conductivity patterns. The remainder of this paper is organized as follows. Sec. \ref{sec:method} provides the theoretical model, including the Hamiltonian and the general procedure calculating the optical conductivity according to the Kubo formula and the Green's function. Sec. \ref{res_1} investigates the numerical evaluations of the optical conductivity in the heterostructure under different Rashba SOC values. With rising Rashba SOC, a second peak appears, and the strengths of both peaks and the frequencies where the peaks arise in the conductivity become higher. This results from the increase of the energy gap and the possibility of electron transition. Sec. \ref{res_2} performs the twist-angle dependence of the conductivity. The chalcogen atoms (S, Se) rather than the metal atoms (Mo, W) influence the conductivity. Sec. \ref{res_3} studies the temperature's effect on the optical conductivity. Besides the Drude peak, another peak occurs in the conductivity due to a joint effect of the thermal and photon transitions. However, in addition to the two peaks, the conductivity becomes lower in other frequency regimes. This is because the thermal excitation restrains the photon excitation. Finally, Sec. \ref{sec:conclu} summarizes the results and presents some concluding remarks.

\section{Method}
\label{sec:method}
To analyze and discuss the optical conductivity of graphene on TMDC, it is necessary to investigate the structure of the energy bands and establish an equation for the electronic Green's function. First, we recall graphene's lattice vectors are expressed as $\mathbf{a}_1=a(1,0)$ and $\mathbf{a}_2=a(1/2,\sqrt{3}/2)$, with $\mathrm{a} = 2.46\ \mathrm{\mathring{A}}$. So the reciprocal lattice vectors $\mathbf{b}_i$ are $\mathbf{b}_1=\frac{4\pi}{\sqrt{3}a}(\sqrt{3}/2,-1/2)$ and $\mathbf{b}_2=\frac{4\pi}{\sqrt{3}a}(0, 1)$ according to $\mathbf{a}_i\cdot\mathbf{b}_j=2\pi\delta_{ij}$ with $i,j=1,2$. Considering the $p_{z}$ orbit vertical to the plane, the Hamiltonian for graphene is written in the effective form around the Fermi level without losing any important physical properties\cite{li2019twist}:

\begin{equation}
\mathcal{H}=\sum_{\xi}\int d\mathbf{k} \Psi^{\dagger} H^{(\xi)}_{\mathrm{eff}}(\mathbf{k}) \Psi
\end{equation}
where $H^{(\xi)}_{\mathrm{eff}}(\mathbf{k})=H_{\mathrm{G}}^{(\xi)}(\mathbf{k})+V_{\mathrm{eff}}^{(\xi)}$. The notation $\xi=\pm 1$ is the valley index representing two different Diarc points in the first Brillouin zone $\mathbf{K}^{(\xi)}=-\xi(2\mathbf{b_1}+\mathbf{b_2})/3$. Within the linear term, $H_{\mathrm{G}}$ is approximated by $H_{\mathrm{G}}^{(\xi)}=\hbar v_{\mathrm{F}} \mathbf{k} \cdot (\xi\sigma_{x}, \sigma_{y})$, where $v_{\mathrm{F}} \sim 10^6\ \mathrm{m/s}$ is the band velocity at the Fermi level of graphene, $\mathbf{k}=(k_{x}, k_{y})$ is the electronic momentum, and $\sigma_{x}$ and $\sigma_{y}$ is the Pauli matrix in the sublattice space. When graphene is placed on the top of the TMDC monolayer, the effective potential $V_{\mathrm{eff}}$ can be gained by integrating out the states of the TMDC layer in the continuum model Hamiltonian\cite{david2019induced}:
\begin{equation}
\label{eq_2}
V_{\mathrm{eff}}=\xi\mathrm{h_{z}}s_{z}+\alpha e^{-i\phi s_{z}/2}(\xi\sigma_{x}s_{y}-\sigma_{y}s_{x})e^{i\phi s_{z}/2}
\end{equation}
where $s_{i}(i=x,y,z)$ is the Pauli matrix in the spin space. Because the spatial mirror symmetry of graphene is broken, the terms $h_z$ and $\alpha$ representing Zeeman energy and Rashba SOC arise. It means the spin axis can be rotated by an angle $\phi$ on the graphene plane\cite{li2019twist,david2019induced}, which contributes nothing to the conductivity, as proved in the following calculation, however. If the eigenvector $\{X, s\}=\{A,\uparrow\},\{B,\uparrow\},\{A,\downarrow\},\{B,\downarrow\}$ is used, 
the effective Hamiltonian can be expressed in an explicit matrix form:
\begin{equation}
H_{\mathrm{eff}}^{(+)}=
\begin{pmatrix}
h_{z} & \hbar v_{\mathrm{F}}(k_{x}-ik_{y}) & 0 & 0\\
\hbar v_{\mathrm{F}}(k_{x}+ik_{y}) & h_{z} & -2i\alpha e^{-i\phi} & 0\\
0 & 2i\alpha e^{i\phi} & -h_{z} & \hbar v_{\mathrm{F}}(k_{x}-ik_{y})\\
0 & 0 & \hbar v_{\mathrm{F}}(k_{x}+ik_{y}) & -h_{z}\\
\end{pmatrix}
\end{equation}

\begin{equation}
H_{\mathrm{eff}}^{(-)}=
\begin{pmatrix}
-h_{z} & \hbar v_{\mathrm{F}}(k_{x}-ik_{y}) & 0 & 2i\alpha e^{-i\phi}\\
\hbar v_{\mathrm{F}}(k_{x}+ik_{y}) & -h_{z} & 0 & 0\\
0 & 0 & h_{z} & \hbar v_{\mathrm{F}}(k_{x}-ik_{y})\\
-2i\alpha e^{i\phi} & 0 & \hbar v_{\mathrm{F}}(k_{x}+ik_{y}) & h_{z}\\
\end{pmatrix}
\end{equation}
Note that if $\phi=0$ and the order of basis is changed to $\{X, s\}=\{B, \downarrow\},\{A, \uparrow\},\{A, \downarrow\},\{B, \uparrow\}$, the Hamiltonian becomes similar to that of the asymmetric bilayer graphene. The $\alpha$ and $h_{z}$ correspond to the interlayer coupling and  the interlayer
asymmetric potential, respectively\cite{nicol2008optical}. Then it is straightforward to obtain the Green's functions through $\hat{G}^{-1}(z)=z\hat{I}-\hat{H}$, and the definitions of $G_{ij}(z)$ are presented in \ref{sec:A}.
The conductivity for finite photon frequency $\Omega$ is calculated through
the standard procedure of using the Kubo formula\cite{mahan2000many} and according to the relation
\begin{equation}
G_{ij}(z)=\int_{-\infty}^{\infty}\frac{d\omega'}{2\pi}\frac{A_{ij}(\omega')}{z-\omega'}
\end{equation}
Green's function can be substituted in the form of  the spectral function representation, whose form is also presented in \ref{sec:A}. By integrating over $\omega$ and $\mathbf{k}$, the conductivity as a function of $\Omega$ is expressed as
\begin{equation}
\sigma_{\alpha \beta}(\Omega)=\frac{N_{f}e^2}{2\Omega}\int_{-\infty}^{\infty}\frac{d\omega}{2\pi}[f(\omega-\mu)-f(\omega+\Omega-\mu)] \times \int\frac{d^{2}k}{{(2\pi)}^2}\mathrm{Tr}[\hat{v_{\alpha}}\hat{A}(\omega+\Omega, \mathbf{k})\hat{v_{\beta}}\hat{A}(\omega, \mathbf{k})]
\end{equation}
where $\alpha, \beta = x, y$, $\hat{v}$ is the velocity operator, $N_f=2$ is the degeneracy because of the two inequivalent Dirac points $\mathbf{K}^{(+)}$ and $\mathbf{K}^{(-)}$, $f(\epsilon)=\frac{1}{e^{{(\epsilon-\mu)}/k_{B}T}+1}$ is the Fermi-Dirac distribution, $k_B$ is the Boltzmann constant and $\mu$ is the chemical potential. It is sufficient to calculate the $v_x$ since only $v_x$ has a contribution to the longitudinal conductivity $\sigma_{xx}$. $v_x$ for both $H_{\mathrm{eff}}^{(+)}$ and $H_{\mathrm{eff}}^{(-)}$ can be written as
\begin{equation}
	v_x=\frac{1}{\hbar}\frac{\partial H_{\mathrm{eff}}}{\partial {k_x}}=v_{\mathrm{F}}
	\begin{pmatrix}
		0 & 1 & 0 & 0\\
		1 & 0 & 0 & 0\\
		0 & 0 & 0 & 1\\
		0 & 0 & 1 & 0\\
	\end{pmatrix}
\end{equation}
By taking the trace and dropping the term which will vanish after being integrated over the whole $\mathbf{k}$ space, we find the $\sigma_{xx}(\Omega)$ can be calculated in the following form:
\begin{equation}
\begin{split}
\sigma_{xx}(\Omega)=\frac{N_{f}e^2}{2\Omega}&\int_{-\infty}^{\infty}\frac{d\omega}{2\pi}[f(\omega-\mu)-f(\omega+\Omega-\mu)] \times \int\frac{d^{2}k}{{(2\pi)}^2}v_{F}^2 \times \\
[+ &A_{22}(\omega+\Omega, \mathbf{k})A_{11}(\omega, \mathbf{k})+ A_{24}(\omega+\Omega, \mathbf{k})A_{31}(\omega, \mathbf{k}) \\
+ &A_{11}(\omega+\Omega, \mathbf{k})A_{22}(\omega, \mathbf{k}) + A_{13}(\omega+\Omega, \mathbf{k})A_{42}(\omega, \mathbf{k}) \\
+ &A_{42}(\omega+\Omega, \mathbf{k})A_{13}(\omega, \mathbf{k}) + A_{44}(\omega+\Omega, \mathbf{k})A_{33}(\omega, \mathbf{k}) \\
+ &A_{31}(\omega+\Omega, \mathbf{k})A_{24}(\omega, \mathbf{k}) + A_{33}(\omega+\Omega, \mathbf{k})A_{44}(\omega, \mathbf{k})]
\end{split}
\end{equation}
%Here the notation $+$ and $-$ correspond to the induces $(xx)$ and $(xy)$, respectively.

\section{Result and Discussion}
\subsection{Longitudinal optical conductivity at different Rashba SOC}
\label{res_1}

\begin{figure}[h]
  \centering
  \begin{subfigure}[h]{0.32\textwidth}
    \includegraphics[width=\textwidth]{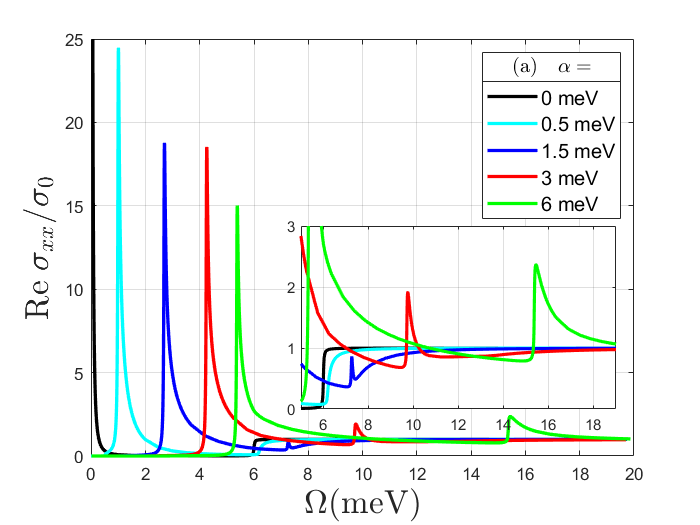}
  \end{subfigure}
  \hfill
  \begin{subfigure}[h]{0.32\textwidth}
    \includegraphics[width=\textwidth]{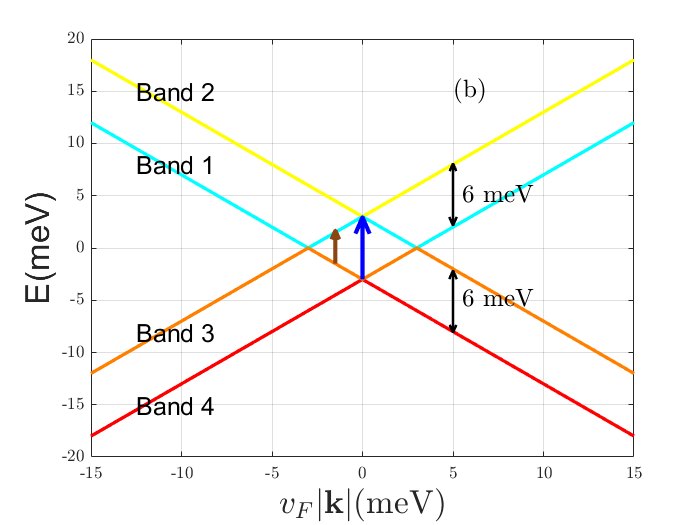}
  \end{subfigure}
  \hfill
  \begin{subfigure}[h]{0.32\textwidth}
    \includegraphics[width=\textwidth]{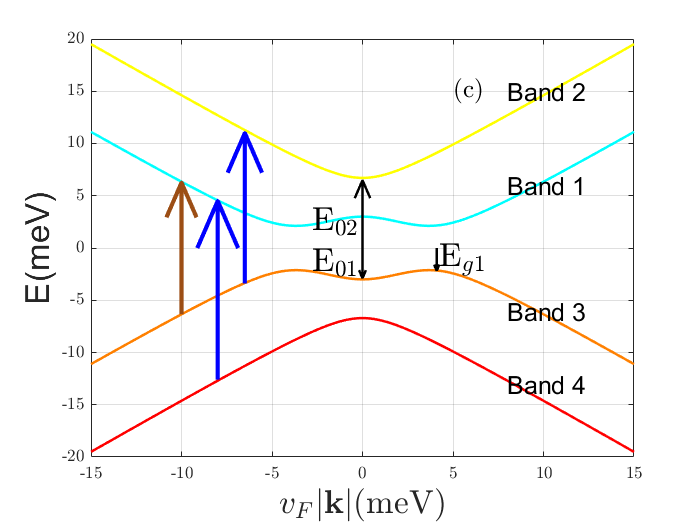}
  \end{subfigure}
  \hfill
  \begin{subfigure}[h]{0.32\textwidth}
    \includegraphics[width=\textwidth]{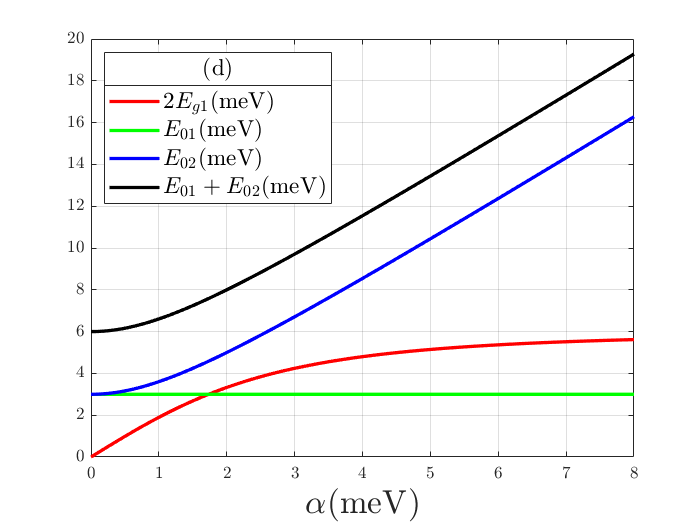}
  \end{subfigure}
  \hfill
  \begin{subfigure}[h]{0.32\textwidth}
    \includegraphics[width=\textwidth]{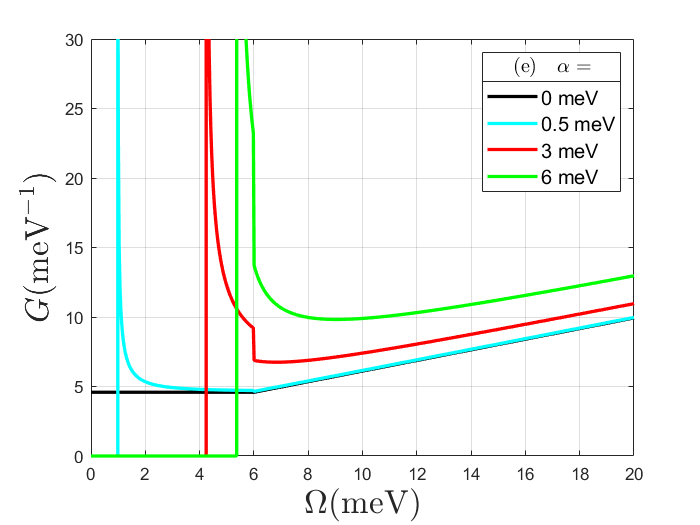}
  \end{subfigure}
  \hfill
  \begin{subfigure}[h]{0.32\textwidth}
    \includegraphics[width=\textwidth]{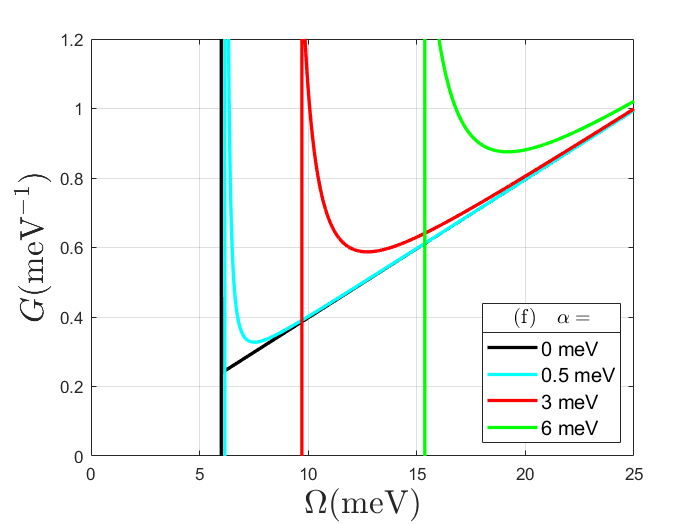}
  \end{subfigure}
  \caption{(a) Real part of longitudinal optical conductivity normalized to $\sigma_0$ as a function of frequency $\Omega$ for five values of the Rashba SOC $\alpha$, as indicated, $T=0$ and $h_z=3\ \mathrm{meV}$. 
  (b-c): The band structure of graphene with spin splitting and Rashba SOC plotted as a function of the momentum $ v_F|\mathbf{k}|$, the spin splitting is fixed at $h_z=3\ \mathrm{meV}$, (b) $\alpha=0$, (c) $\alpha=3\ \mathrm{meV}$. 
  (d) Energy scales as a function of Rashba SOC, $h_z=3\ \mathrm{meV}$
  (e-f): The DOS for band (e)$3\rightarrow1$ and (f)$3\rightarrow2$ and $4\rightarrow1$ are plotted as a function of the band energy, whose notation remains consistent with those in (a).}
  \label{fig:energy}
\end{figure}

In Fig. \ref{fig:energy} (a), the real part of the longitudinal optical conductivity, denoted as $\sigma_{xx}(\Omega)$, is plotted as a function of the photon frequency $\Omega$. For this particular system, the following relation holds $\sigma_{xx}(\Omega)=\sigma_{yy}(\Omega)$. The component 
of the conductivity is normalized by the conductance unit
$\sigma_0=e^2/4\hbar$. In our numerical analysis, to simulate the impurity scattering, the delta functions in the spectral functions are replaced by the Lorentz distribution, which is expressed as $\delta(x)=(\eta/\pi)/(x^2+\eta^2)$. Here, $\eta$ is a boarding parameter representing the effective transport scattering rate. And in the subsequent procedure, we fix $\eta=0.01\mathrm{meV}$ following Ref. \cite{alidoust2022controllable}. For simplicity, $\hbar=1$ is set.

As the photon frequency increases, the longitudinal conductivity approaches the conductance unit $\sigma_0$, which is in accordance with Ref. \cite{falkovsky2007space}. It is evident that when $\alpha=0$, there is a Drude response at the frequency $\Omega=0$. The conductivity shows a sharp step jumping to the conductance unit $\sigma_0$ when $\Omega=6\ \mathrm{meV}$. Note that although the conductivity is extremely low between $\Omega\approx2\ \mathrm{meV}$ and $6\ \mathrm{meV}$, it does not vanish. 
%While the transverse conductivity is always zero in the frequency regime.
When Rashba SOC is increased to a nonzero value, e.g., $\alpha=0.5\ \mathrm{meV}$, the longitudinal conductivity shows a semiconductor behavior at zero frequency. 
Besides, It has a drastic peak at $\Omega\approx1\ \mathrm{meV}$, and the sharp step at $\Omega=6\ \mathrm{meV}$ moves right slightly and becomes flatter. A second peak appears upon increasing $\alpha$ further from $1.5$ to $6\ \mathrm{meV}$.
The frequencies where the two peaks appear rise with increasing  $\alpha$. 
When $\alpha$ is $1.5,3,6\ \mathrm{meV}$, the frequency values are $2.7$, $4.2$, $5.4\ \mathrm{meV}$ for the first peak and $7.2$, $9.7$, $15.4\ \mathrm{meV}$ for the second peak, respectively. Besides, the two peaks show an enhancement in width and height when $\alpha$ increases. The first peak is higher and wider than the second peak. In the low-frequency regime, the extent of the zero-conductance region can be extended by increasing the magnitude of Rashba SOC until it reaches a critical threshold.

To uncover the physical origins of the optical conductivity results above, the energy bands and DOS for different $\alpha$ is plotted in Fig. \ref{fig:energy} (b-f).  For consistency, the parameters are the same as those in Fig. \ref{fig:energy} (a). The eigenvalues are written as:
\begin{equation}
E_{1,2,3,4} = \pm\sqrt{h_z^2+2\alpha^2+|\mathbf{k}|^2v_F^2\pm2\sqrt{\alpha^4+|\mathbf{k}|^2v_F^2h_z^2+|\mathbf{k}|^2v_F^2\alpha^2}}
\end{equation}
%and the analytical form of density of states(DOS) for different bands are presented in Appendix \ref{sec:C}. 
In band 1, there is a local maximum at zero momentum $\mathbf{k}=0$ with energy $E_{01}=h_z$ and two minima at $\mathbf{k}=\pm h_z\sqrt{(h_z^2+2\alpha^2)/(h_z^2+\alpha^2)}$ with energy  $E_{g1}=h_z\alpha/\sqrt{h_z^2+\alpha^2}$. The lowest energy for band 2 occurs at $\mathbf{k}=0$ with energy $E_{02}=\sqrt{4\alpha^2+h_z^2}$. The three energy scales are the key to understanding the physics of the conductivity.

The DOS identified here is $G=dN/d\Omega =2\pi |\mathbf{k}|\cdot \frac{dN}{d|\mathbf{k}|}\cdot \frac{d|\mathbf{k}|}{d\Omega}$ to describe the possibility of electron interband transition at different photon values. The transition between different bands holds different relationships. For band $3\rightarrow1$, $\Omega=2E_1(|\mathbf{k}|)$ and for band $3\rightarrow2$/$4\rightarrow1$, $\Omega=E_1(|\mathbf{k}|)+E_2(|\mathbf{k}|)$. The DOS can be written as:
\begin{equation}
\begin{aligned}
&G_{3\rightarrow1}(\Omega)=[P_{+}(\Omega)-P_{-}(\Omega)\Theta(2E_{01}-\Omega)]\Theta(\Omega-2E_{g1}) \\
&P_{\pm}(\Omega)=\frac{1}{\pi\hbar^2v_F^2}\frac{\Omega}{1-(h_z^2+\alpha^2)/F^{\pm}(\Omega)} \\
&F^{\pm}(\Omega)=\sqrt{\alpha^4+(h_z^2+\alpha^2)[(\Omega/2)^2+h_z^2\pm2\sqrt{(\Omega/2)^2(h_z^2+\alpha^2)-h_z^2\alpha^2}]}
\end{aligned}
\end{equation}
and
\begin{equation}
\begin{aligned}
&G_{3\rightarrow2,4\rightarrow1}(\Omega)=\frac{1}{2\pi}\frac{1}{\sum_{i=1,2}Q_i}\Theta(\Omega-(E_{01}+E_{02})) \\
&Q_{1,2}={(1\pm{h_z^2+\alpha^2/\sqrt{\alpha^4+(h_z^2+\alpha^2)|\mathbf{k}|^2(\Omega)}}})/{E_{1,2}(|\mathbf{k}|^2(\Omega))} \\
&|\mathbf{k}|^2(\Omega)=\frac{\Omega^4+16\alpha^4-4\Omega^2(h_z^2+2\alpha^2)}{4\Omega^2+16(h_z^2+\alpha^2)}
\end{aligned}
\end{equation}
The indices 1,2 correspond to the signs $+$ and $-$, respectively.

Recall that we use the Lorentz distribution to replace the delta function in this paper, so the peaks in the optical conductivity are boarded. When $\alpha=0$, there is no energy gap between the energy bands, which means graphene is like metal under this condition. As a result, the Direct Current(DC) longitudinal conductivity shows a metallic Drude response. There is also a possibility of electron transition  band $3\rightarrow1$ with increasing frequency, as indicated by the brown arrow in Fig. \ref{fig:energy} (b). However, the corresponding DOS stays small, thus explaining why conductivity in the regime is small but not zero. When $\Omega=6\ \mathrm{meV}$, the transition 
 possibility is enhanced  because other photon transitions: band $3\rightarrow2$ and $4\rightarrow1$ are opened, as indicated by the dark blue arrows in Fig. \ref{fig:energy} (b).
 And because of the linear dispersion, the photon energy $\Omega=6\ \mathrm{meV}$ can promote electron excitation in the regime $v_F |\mathbf{k}|<3\ \mathrm{meV}$. As a result, there is a sharp step in the conductivity at $\Omega=6\ \mathrm{meV}$.
 When $\alpha=0.5\ \mathrm{meV}$, an energy band opens and photon energy $2E_{g1}\approx1\ \mathrm{meV}$ is required for interband transition band $3\rightarrow1$, accounting for the peak appearing in the conductivity in Fig. \ref{fig:energy} (a). And its linear dispersion has been destructed slightly and excitation energy has been changed from a value $\Omega=6\ \mathrm{meV}$ to a range between $\Omega\approx6.2\ \mathrm{meV}$ 
 and $7.2\ \mathrm{meV}$. As a result, the conductivity step around $\Omega=6\ \mathrm{meV}$ is boarded.
 
 Upon increasing $\alpha$ from $1.5$ to $6\ \mathrm{meV}$, the first peak moves right continuously. This still can be explained by the increase of the energy gap $2E_{g1}$. The van Hove singularity of band 1 has been enhanced enormously as $\alpha$ increases, which means more electrons can be excited from band 3 to band 1(the brown arrow in Fig. \ref{fig:energy} (e)). It contributes to the enhancement of the first conductivity peak. 
 The second peak results from the electron transition band $3\rightarrow2$ and $4\rightarrow1$(the dark blue arrows in Fig. \ref{fig:energy} (c)).
 %The appearance of the second peak and the increasing strengths of the two peaks can be explained by the corresponding DOS. When $\alpha$ increases, the DOS in Fig. \ref{fig:energy} (f) also increases, just like the behavior of the van Hove singularity of band 1. 
 When $\alpha$ increases, the DOS peak becomes higher. This means there are more available states for interband transitions, thus explaining the strength enhancement of the second conductivity peak.
 However, the DOS for $\alpha=0.5\ \mathrm{meV}$ is so small that the peak is too weak and cannot be observed. 
 The width of peaks in Fig. \ref{fig:energy} (f) can account for the  sharpness of the conductivity step in Fig. \ref{fig:energy} (a). When the slope of DOS approaches the constant $1/(8\pi)$, the conductivity approaches the conductance unit $\sigma_0$. With increasing $\alpha$, there is an enhancement in the width of DOS peak in Fig. \ref{fig:energy} (f), thus causing a flatter step in \ref{fig:energy} (a). The relative conductivity strengths of the two peaks can be understood by the relative DOS strengths. The zero conductivity in the low-frequency regime results from the prohibition of the interband transition.

 %In Frame (c), it is clear after the rapid decrease of DOS of band 1 around the van Hove singularity, it still slowly increases with increasing energy $E_{01}$. This can account for the slow conductivity increase after the second peak. Actually, it is also the reason of the increase of the conductivity at $\alpha=0$ and $0.5\ \mathrm{meV}$.

\begin{figure}[h]
\centering
  \begin{subfigure}[h]{0.49\textwidth}
    \includegraphics[width=\textwidth]{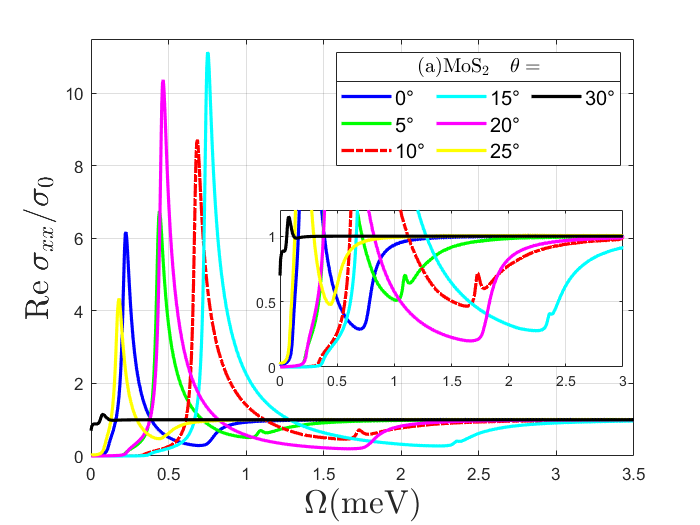}
  \end{subfigure}
  \hfill
  \begin{subfigure}[h]{0.49\textwidth}
    \includegraphics[width=\textwidth]{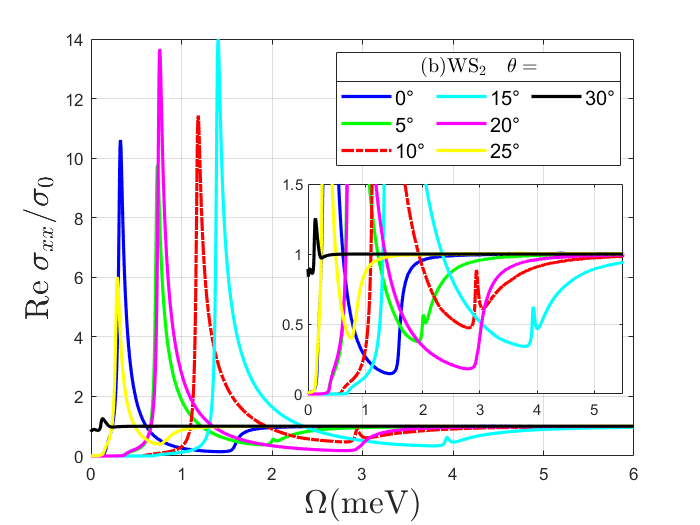}
  \end{subfigure}
  \hfill
  \begin{subfigure}[h]{0.49\textwidth}
    \includegraphics[width=\textwidth]{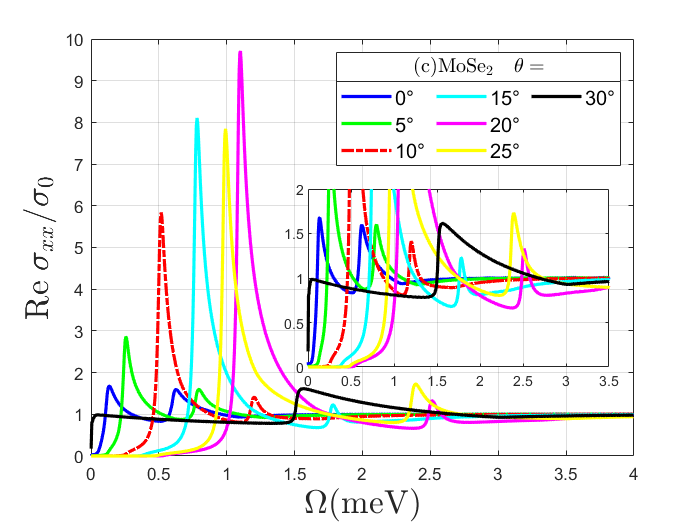}
  \end{subfigure}
  \hfill
  \begin{subfigure}[h]{0.49\textwidth}
    \includegraphics[width=\textwidth]{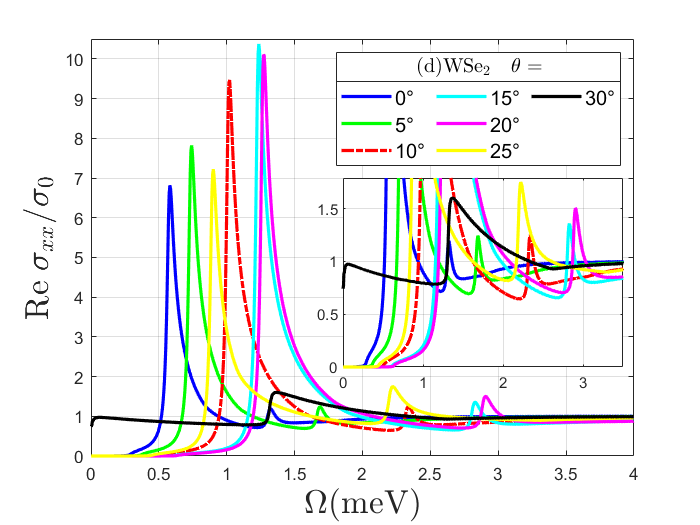}
  \end{subfigure}

\caption{The real part of frequency-dependent longitudinal optical conductivity $\sigma_{xx}(\Omega)$ normalized to $\sigma_0$ for seven values of twisted-angle as indicated when graphene is placed and twisted on TMDC at $T=\mathrm{0}$: (a)$\mathrm{MoS_2}$, (b)$\mathrm{WS_2}$, (c)$\mathrm{MoSe_2}$, (d)$\mathrm{WSe_2}$.}
\label{fig:mainfig}
\end{figure}

\subsection{Longitudinal optical conductivity at different twisted angles}
\label{res_2}
In Fig. \ref{fig:mainfig}, the longitudinal conductivity as a function of frequency $\omega$ under zero temperature in the twisted heterostructure is plotted. The values of $\alpha$ and $h_z$ under different twisted angles $\theta=0, 5^\circ, 10^\circ, 15^\circ, 20^\circ, 25^\circ, 30^\circ$ are extracted from Ref. \cite{li2019twist} and presented in \ref{sec:B}.

For the four TMDCs, as the angle increases from $0$ to $30^{\circ}$, the frequency where the first peak arises increases then decreases, and the peak reaches the highest and widest point between $\theta=15^{\circ}$ and $25^{\circ}$. 
This is because, in this twist-angle regime, the values for $\alpha$ and $h_z $ become much larger than that of $\theta=0$ and $\theta=30^{\circ}$.
Besides, the upper two subfigures ($\mathrm{MoS_2,WS_2}$) are distinctly different from the lower two ones ($\mathrm{MoSe_2,WSe_2}$). Namely, the chalcogen atoms ($\mathrm{S,Se}$) rather than the metal atoms ($\mathrm{Mo,W}$) have a prominent influence on the conductivity.

An obvious difference between the graphene conductivity on the two types of TMDCs occurs at $\theta=30^{\circ}$, where Rashba SOC $\alpha$ vanishes because of $C_6$ symmetry. The high values for $h_z$ cause the peaks of the conductivity of graphene on $\mathrm{MoSe_2}$ and $\mathrm{WSe_2}$ to be flatter and more obvious than those on $\mathrm{MoS_2}$ and $\mathrm{WS_2}$. They also cause another interesting phenomenon: the second peak is much stronger than the first peak in the conditions of $\mathrm{MoSe_2}$ and $\mathrm{WSe_2}$. However, for $\mathrm{MoS_2}$ and $\mathrm{WS_2}$, when $\theta=30^\circ$, the peaks are much weaker than those at other angles because of the low values for $h_z$.

For the conductivity of graphene placed on $\mathrm{MoS_2}$ and $\mathrm{WS_2}$, the second peak damps out when $\theta\rightarrow0$ or $\theta>20^{\circ}$ except $\theta=30^{\circ}$, although the second peak between $\theta=5^\circ$ and $20^\circ$ is negligible compared to the first peak. By contrast, when considering that on $\mathrm{MoSe_2}$ and $\mathrm{WSe_2}$, a different pattern occurs. Because $\alpha$ is comparable to $h_z$ in the conditions of $\mathrm{MoSe_2}$ and $\mathrm{WSe_2}$, at all angles except $30^{\circ}$, the second peak always exist although it is still weaker than the first peak. While at $\theta=0$, the graphene conductivity on $\mathrm{MoSe_2}$ and $\mathrm{WSe_2}$ shows a different response to the optical frequency. The height and width of the two peaks for $\mathrm{MoSe_2}$ are almost the same. 
By contrast, the first peak for $\mathrm{WSe_2}$ is much stronger than the second peak. This is because, for $\mathrm{WSe_2}$,  $h_z$ is much smaller than $\alpha$ at this angle.

\subsection{Longitudinal optical conductivity at different temperature}
\label{res_3}
\begin{figure}[h] % 可选参数用于指定图片的位置
  \centering
  \begin{subfigure}[b]{0.32\textwidth}
    \includegraphics[width=\textwidth]{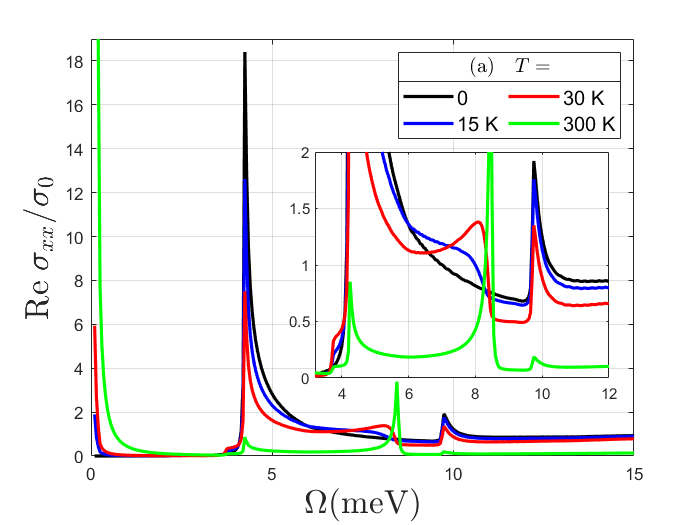}
  \end{subfigure}
  \hfill
  \begin{subfigure}[b]{0.32\textwidth}
    \includegraphics[width=\textwidth]{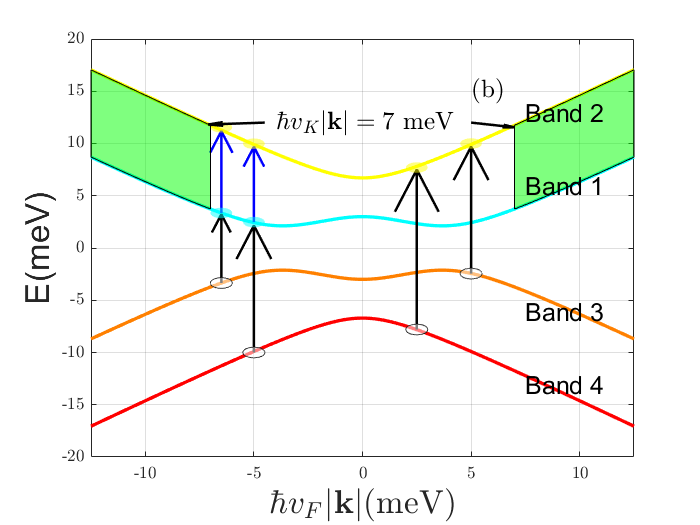}
  \end{subfigure}
  \hfill
  \begin{subfigure}[b]{0.32\textwidth}
    \includegraphics[width=\textwidth]{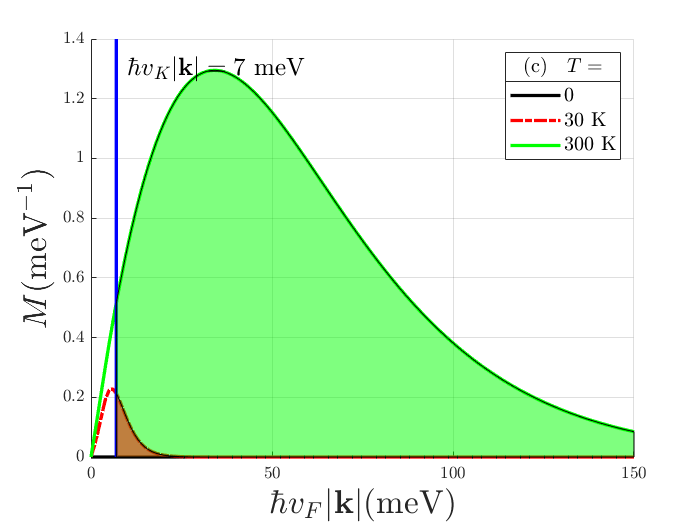}
  \end{subfigure}
  \caption{(a) Real part of frequency-dependent longitudinal optical conductivity $\sigma_{xx}(\Omega)$ normalized to $\sigma_0$ for four values of temperature as indicated, $\alpha$ and $h_z$ is both $3\mathrm{meV}$. (b) $G=f(E_1(\mathbf{k})) \cdot dN/|d\mathbf{k}|$ as a function of $\hbar v_F |\mathbf{k}|$ to describe the number of electrons in band 1.}
  \label{fig:example3}
\end{figure}
In Fig. \ref{fig:example3} (a), the behaviors of the longitudinal conductivity at different temperatures $T=0,15,30,300\ \mathrm{K}$ are shown as a function of frequency. Here, we set $\alpha=3\ \mathrm{meV}$ and $h_z=3\ \mathrm{meV}$. Note that when frequency $\Omega\rightarrow\infty$, the conductivity still approaches $\sigma_0$. However, plotting the conductivity from $\Omega=0$ to 15 $\mathrm{meV}$ is sufficient, which contains all the interesting physical results.

Besides the physical results described above, three other interesting phenomena occur. First of all, like the behavior of general semiconductors, as long as $T\neq 0$, there is a Drude weight whose strength increases with temperature when $\Omega\rightarrow0$. In addition, a band arises between $\Omega=3.7\ \mathrm{meV}$ and $8.5\ \mathrm{meV}$. A maximum is around $\Omega\approx8.5\ \mathrm{meV}$, gradually evolving into a sharp peak with rising temperature. The frequency where the maximum occurs becomes higher slightly when $T$ rises from $30\ \mathrm{K}$ to $300\ \mathrm{K}$. In addition, in other conductivity ranges, the values for the conductivity decrease dramatically as the temperature $T$ increases, which is in accordance with Ref. \cite{falkovsky2007space}.  

%First, as long as $T\neq 0$, there is a Drude weight increasing with temperature when $\omega\rightarrow0$. This is because that more electrons are thermally excited from the valence band to the conduction band as temperature increases. By contrast, there is a different trend occurring between $\omega=11\mathrm{meV}$ and $\omega=15\mathrm{meV}$, which is consistent with Ref. \cite{falkovsky2007space}. However, an interesting phenomenon is that a band arises that between $\omega=3.7\mathrm{meV}$ and $\omega=8.5\mathrm{meV}$. There is a maximum around $\omega\approx8.5\mathrm{meV}$, which gradually elvoves a sharp peak with increasing temperature. And the frequency where the maximum occurs move downwards slightly when $\mathrm{T}$ rises from $\mathrm{30K}$ to $\mathrm{300K}$. 

Gaining further insight into the physical origins of the optical conductivity results above is necessary. To begin with, one can estimate the influence of temperature. When $T=30\ \mathrm{K}$, the thermal voltage can be calculated to be $2.6\ \mathrm{meV}$ using the definition $\mathrm{V_T}=k_{B}T/e$, which is close to the values of $\alpha$ and $h_z$ used here. As a result, it is straightforward to conclude that the temperature impact is obvious when $T=30\ \mathrm{K}$. 

%Electrons and holes contribute to the optical conductivity by both thermal excitation and photonic excitation. 

When $T\neq 0$, there are electrons excited thermally from the valence band(3,4) to the conduction band(1,2), as indicated by the black arrows in Fig. \ref{fig:example3} (b). Accordingly, electrons in the conduction bands and holes in the valence bands contribute to the nonzero conductivity. When $T$ increases, more electrons are excited into the conduction bands and the optical conductivity increases. Note that this depends completely on the thermal excitation because zero photon excitation contributes nothing to the Drude conductivity.

The increasing number of electrons in band 1 provides more possibility to be excited into band 2, as indicated by the dark blue arrow in Fig. \ref{fig:example3} (b). This causes the the conductivity band mentioned above from $3.7\ \mathrm{meV}$ to $8.5\ \mathrm{meV}$. It is a joint effect of the two types of excitation because the conductivity band results from the optical transition band $1\rightarrow2$ and the electrons in band 1 are thermally excited from valence bands. Here we give a semi-quantitative explanation in Fig. \ref{fig:example3} (c). Considering temperature, we plot $M=f(E_1(\mathbf{k})) \cdot dN/|d\mathbf{k}|$ as a function of momentum $v_F |\mathbf{k}|$ to describe the electrons excited thermally in band 1. When $T=0$, there are no electrons in band 1. As the temperature rises from $30\ \mathrm{K}$ to $300\ \mathrm{K}$, $G$ becomes much lager, thus explaining why the conductivity is much larger. When $\hbar v_F |\mathbf{k}|>7\mathrm{meV}$, the energy difference between the two conduction energy bands approaches a constant $\Delta E\approx8.5\ \mathrm{meV}$, which means the electrons in band 1 can be excited by the certain photon energy. When temperature increases, more electrons will occur in the momentum range when $\hbar v_F |\mathbf{k}|>7\mathrm{meV}$, thus increasing conductivity at $\Omega\approx8.5\ \mathrm{meV}$.
And the maximum of $G$ has moved right with increasing temperature. This is the reason why the frequency where the conductivity maximum arises at $T=300\ \mathrm{K}$ around $8.5\ \mathrm{meV}$ is slightly higher than that at $T=30\ \mathrm{K}$.

However, thermal excitation and photonic excitation can also be mutually inhibitory. The existence of the electrons in the conduction bands restricted the photon excitation of electrons from bands with lower energy. This accounts for  why except for the two peaks mentioned above, the conductivity becomes lower in the other conductivity ranges.  

\section{Conclusion}
\label{sec:conclu}
To conclude, We have examined the optical graphene conductivity in the heterostructure by varying the values of Rashba SOC. As Rashba SOC increases, we have observed the emergence of a second peak in the conductivity. The strengths of the two peaks and the frequencies at which these peaks occur also increase. This phenomenon can be attributed to the increase in the energy gap and the possibility of electron transition. We have also investigated the impact of the twist-angles on the conductivity. Notably, the chalcogen atoms (S, Se) had a more significant influence on conductivity than the metal atoms (Mo, W). Furthermore, the dependence of the optical conductivity on temperature is studied. Besides the Drude peak resulting from the thermal excitation, another peak appears in the conductivity. This is because of the joint effect of the thermal transition and the photon transition. However, in other frequency ranges, the conductivity decreased due to thermal excitation suppressing photon excitation. In the future, the thermal vibrations of the carbon atoms are also supposed to be added to the total conductivity when the temperature is high enough. The intraband peak in the optical conductivity can also be taken into account\cite{huang2023nonlocal}. This can deepen our understanding of graphene conductance and help us to achieve its regulation and application by changing external conditions.

\begin{Declaration of competing interest}
The authors declare that they have no known competing financial interests or personal relationships that could have appeared to influence the work reported in this paper.
\end{Declaration of competing interest}

\begin{acknowledgment}
Ruoyang Cui thanks Zhe Feng from HHU for continued support and Yechun Ding from XJTU for useful discussion.
\end{acknowledgment}

\appendix
\section{Derivation of Effective Potential $V_{\mathrm{eff}}$}
\label{sec:C}
In this appendix, the specific procedure for deriving the effective potential in eqn. \ref{eq_2} is presented. The procedure has been discussed in ref. \cite{li2019twist} thoroughly. The Hamiltonian is spanned by the Bloch bases $\ket{\mathbf{k},X,s}$ and $\ket{\mathbf{\widetilde{k}},\widetilde{X},\widetilde{s}}$:
\begin{equation}
	\begin{aligned}
		&\ket{\mathbf{k},X,s}=\frac{1}{\sqrt{N}}\sum_{\mathbf{R}_X}e^{i\mathbf{k\cdot\mathbf{R}_X}}\ket{\mathbf{R}_X,s} \: (\mathrm{graphene})\\
		&\ket{\mathbf{\widetilde{k}},\widetilde{X},\widetilde{s}}=\frac{1}{\sqrt{\widetilde{N}}}\sum_{\mathbf{R}_{\widetilde{X}}}e^{i\mathbf{\widetilde{k}\cdot\mathbf{R}_{\widetilde{X}}}}\ket{\mathbf{R}_{\widetilde{X}},\widetilde{s}} \: (\mathrm{TMDC}) \\
	\end{aligned}
\end{equation}
with 
\begin{equation}
	\begin{aligned}
		&\mathbf{R}_X=n_1\mathbf{a}_1+n_2\mathbf{a}_2+\mathbf{\tau}_X\\
		&\mathbf{R}_{\widetilde{X}}=\widetilde{n}_1\widetilde{\mathbf{a}}_1+\widetilde{n}_2\widetilde{\mathbf{a}}_2+\mathbf{\tau}_{\widetilde{X}}
	\end{aligned}
\end{equation}
where $\mathbf{k}$ and $\mathbf{\widetilde{k}}$ are Bloch wave vectors, $N$ and $\widetilde{N}$ are the number of unit cells contained in the heterostructure unit cell, $s$ and $\widetilde{s}$ are spin indexes, $\ket{\mathbf{R}_X,s}$ and $\ket{\mathbf{R}_{\widetilde{X}},\widetilde{s}}$ represent the Wannier bases, $n_1,n_2$ and  $\widetilde{n}_1,\widetilde{n}_2$ are integers, $\tau_X$ and $\tau_{\widetilde{X}}$ are the sublattice positions inside the unit cell. Then the Hamiltonian around the Dirac point for the graphene/TMDC heterostructure can be represented as

\begin{equation}
	H=
	\begin{pmatrix}
		H_G^{(\xi)}(\mathbf{k}) & T^{(\xi),1} & T^{(\xi),2} & T^{(\xi),3}\\
		T^{\dagger,(\xi),1} & H_T^{(\xi),1}(\mathbf{\widetilde{k}}_1) & 0 & 0\\
		T^{\dagger,(\xi),2} & 0 & H_T^{(\xi),2}(\mathbf{\widetilde{k}}_2) & 0\\
		T^{\dagger,(\xi),3} & 0 & 0 & H_T^{(\xi),3}(\mathbf{\widetilde{k}}_3)\\
	\end{pmatrix}
\end{equation}
where $T^{(\xi),j}=\bra{\mathbf{k},X,s}H_{\mathrm{int}}\ket{\mathbf{\widetilde{k}}_j,\widetilde{X},\widetilde{s}} \ \ (j=1,2,3)$ means the interlayer interaction. For $\mathbf{k}=\mathbf{K}_{\xi}$, the dominant contribution mainly comes from three reciprocal points in the TMDC layer, $\mathbf{\widetilde{k}}_1=\mathbf{K}_{\xi}+\xi\mathbf{\widetilde{b}}_1$, $\mathbf{\widetilde{k}}_2=\mathbf{K}_{\xi}+\xi(\mathbf{b}_1+\mathbf{\widetilde{b}}_2)$,
$\mathbf{\widetilde{k}}_3=\mathbf{K}_{\xi}+\xi(\mathbf{b}_1+\mathbf{b}_2-\mathbf{\widetilde{b}}_1-\mathbf{\widetilde{b}}_2)$ where $\mathbf{b}_1$, $\mathbf{b}_2$ and $\mathbf{\widetilde{b}}_1$, $\mathbf{\widetilde{b}}_2$ are unit reciprocal lattice vectors for graphene and TMDCs, respectively. The influence of other $\mathbf{\widetilde{k}}$'s is negligibly small. Accordingly, it is sufficient to restrict the Hamiltonian with these three interactions. 
	
In order to gain an effective Hamiltonian, the Schrieffer-Wolff transformation is applied\cite{schrieffer1966relation,bravyi2011schrieffer}:  $H^{(\xi)}_{\mathrm{eff}}(\mathbf{k})=H_{\mathrm{G}}^{(\xi)}(\mathbf{k})+V_{\mathrm{eff}}^{(\xi)}(\mathbf{k})$. The form of $H_{\mathrm{G}}^{(\xi)}(\mathbf{k})$ is showed in Sec. \ref{sec:method}, and that of $V_{\mathrm{eff}}^{(\xi)}(\mathbf{k})$ is written as
\begin{equation}
	[V^{(\xi)}_{\mathrm{eff}}(\mathbf{k})]_{X's',Xs}=\sum_{\widetilde{n},\mathbf{\widetilde{k}}}\frac{\bra{\mathbf{k},X',s'}H_{\mathrm{int}}\ket{\widetilde{n},\mathbf{\widetilde{k}}}\bra{\widetilde{n},\mathbf{\widetilde{k}}}H_{\mathrm{int}}\ket{\mathbf{k},X,s}}{E_G-E_{\widetilde{n},\mathbf{\widetilde{k}}}}
\end{equation}
where $E_G$ is energy of the Dirac point and $E_{\widetilde{n},\mathbf{\widetilde{k}}}$ and $\ket{\widetilde{n},\mathbf{\widetilde{k}}}$ are the eigenvalue and eigenstate of $H_T^{(\xi),1}(\mathbf{k})$, with the band index $\widetilde{n}$ (including the spin
degree of freedom) and the Bloch vector $\mathbf{\widetilde{n}}$. After the summation, 
the zeroth order $V^{(\xi)}_{\mathrm{eff}}(\mathbf{k})=V^{(\xi)}_{\mathrm{eff}}$ is taken to gain the proximity SOC:
\begin{equation}
	V_{\mathrm{eff}}^{(\xi)}=\xi\mathrm{h_{z}}s_{z}+\alpha e^{-i\phi s_{z}/2}(\xi\sigma_{x}s_{y}-\sigma_{y}s_{x})e^{i\phi s_{z}/2}
\end{equation}

\section{Green's Functions And Spectral Functions}
\label{sec:A}

In this appendix, we present the Green's functions and spectral functions for the Hamiltonian above. As $G_{44}^{(+)}(\lambda)=G_{11}^{(+)}(-\lambda)$, $G_{33}^{(+)}(\lambda)=G_{22}^{(+)}(-\lambda)$, and $G_{24}^{(+)}(\lambda)=G_{13}^{(+)}(-\lambda)$, It is sufficient to present the following three elements and the other terms occurring in Equation can be calculated using the Hermitian property of the Hamiltonian. 
\begin{equation}
\Omega G_{11}^{(+)}(z)=-4(h_{z}+z)\alpha^2-(h_{z}-z)(h_{z}-\mathbf{k}+z)(h_{z}+\mathbf{k}+z)
\end{equation}

\begin{equation}
\Omega G_{13}^{(+)}=2ie^{-i\phi}\alpha(k_x-ik_y)(z-h_z)
\end{equation}

\begin{equation}
\Omega G_{22}^{(+)}=(h_z-\mathbf{k}+z)(h_z+\mathbf{k}+z)(h_{z}-z)
\end{equation}
And their corresponding spectral functions are
\begin{equation}
\begin{split}    
\frac{A_{11}^{(+)}(\omega)}{2\pi}=&\frac{1}{4}[\delta(\omega+\mathcal{K})+\delta(\omega-\mathcal{K})+\delta(\omega+\mathcal{L})+\delta(\omega-\mathcal{L})]\\
- &\frac{2\alpha^2}{\mathcal{D}}[\delta(\omega+\mathcal{K})+\delta(\omega-\mathcal{K})-\delta(\omega+\mathcal{L})-\delta(\omega-\mathcal{L})]\\
+ &\frac{h_z}{\mathcal{K}\mathcal{D}}(\frac{1}{4}\mathcal{D}+2{|\mathbf{k}|}^2-2\alpha^2)[\delta(\omega+\mathcal{K})-\delta(\omega-\mathcal{K})]\\
+ &\frac{h_z}{\mathcal{L}\mathcal{D}}(\frac{1}{4}\mathcal{D}-2{|\mathbf{k}|}^2+2\alpha^2)[\delta(\omega+\mathcal{L})-\delta(\omega-\mathcal{L})]
\end{split}
\end{equation}

\begin{equation}
\begin{split}    
\frac{A_{13}^{(+)}(\omega)}{2\pi}=2ie^{-i\phi}\alpha(k_x-ik_y)\times\{&-\frac{1}{\mathcal{D}}[\delta(\omega+\mathcal{K})+\delta(\omega-\mathcal{K})-\delta(\omega+\mathcal{L})-\delta(\omega-\mathcal{L})]\\
-&\frac{h_z}{\mathcal{K}\mathcal{D}}[\delta(\omega+\mathcal{K})-\delta(\omega-\mathcal{K})]\\
+&\frac{h_z}{\mathcal{L}\mathcal{D}}[\delta(\omega+\mathcal{L})-\delta(\omega-\mathcal{L})]\}
\end{split}
\end{equation}

\begin{equation}
\begin{split}    
\frac{A_{22}^{(+)}(\omega)}{2\pi}=&\frac{1}{4}[\delta(\omega+\mathcal{K})+\delta(\omega-\mathcal{K})+\delta(\omega+\mathcal{L})+\delta(\omega-\mathcal{L})]\\
+ &\frac{2\alpha^2}{\mathcal{D}}[\delta(\omega+\mathcal{K})+\delta(\omega-\mathcal{K})-\delta(\omega+\mathcal{L})-\delta(\omega-\mathcal{L})]\\
+ &\frac{h_z}{\mathcal{K}\mathcal{D}}(\frac{1}{4}\mathcal{D}+2{|\mathbf{k}|}^2+2\alpha^2)[\delta(\omega+\mathcal{K})-\delta(\omega-\mathcal{K})]\\
+ &\frac{h_z}{\mathcal{L}\mathcal{D}}(\frac{1}{4}\mathcal{D}-2{|\mathbf{k}|}^2-2\alpha^2)[\delta(\omega+\mathcal{L})-\delta(\omega-\mathcal{L})]
\end{split}
\end{equation}
where
\begin{equation}
\mathcal{C}=4h_z^2+8\alpha^2+4|\mathbf{k}|^2
\end{equation}
\begin{equation}
\mathcal{D}=8\sqrt{\alpha^4+h_z^2|\mathbf{k}|^2+\alpha^2\ |\mathbf{k}|^2}
\end{equation}
\begin{equation}
\mathcal{K}=\frac{1}{2}\sqrt{\mathcal{C}+\mathcal{D}}
\end{equation}
\begin{equation}
\mathcal{L}=\frac{1}{2}\sqrt{\mathcal{C}-\mathcal{D}}
\end{equation}
Besides, since $G_{11}^{(+)}=G_{33}^{(-)}$, $G_{13}^{(+)}=G_{13}^{(-)}$, $G_{22}^{(+)}=G_{44}^{(-)}$, $G_{24}^{(+)}=G_{24}^{(-)}$, $G_{33}^{(+)}=G_{11}^{(-)}$ and $G_{44}^{(+)}=G_{22}^{(-)}$, the degeneracy $N_{f}=2$ and it is enough that only the Dirac point $\xi=+1$ is discussed.
\section{Spin splitting and Rashba SOC on graphene implied by TMDC}
\label{sec:B}

The data of 2$h_z$ and 2$\alpha$ under different twisted angles $\theta=0, 5^\circ, 10^\circ, 15^\circ, 20^\circ, 25^\circ, 30^\circ$ are extracted from Ref. \cite{li2019twist}, which uses the tight-binding model and the perturbational approach.

\begin{table}[h]
\centering
\caption{2$h_z$ at different twist-angles in different graphene/TMDC heterostructures}
\begin{tabular}{cccccccc}
\hline
 2$h_z(\mathrm{meV})$& 0\textdegree & 5\textdegree & 10\textdegree & 15\textdegree & 20\textdegree & 25\textdegree & 30\textdegree \\
\hline
$\mathrm{MoS_2}$  & 0.61 & 0.75 & 1.20 & 1.83 & 1.45 & 0.35 & 0 \\
\hline
$\mathrm{WS_2}$  & 1.51 & 1.69 & 2.29 & 3.32 & 2.74 & 0.69 & 0 \\
\hline
$\mathrm{MoSe_2}$  & 0.12 & 0.27 & 0.66 & 1.18 & 1.66 & 1.22 & 0 \\
\hline
$\mathrm{WSe_2}$  & 0.85 & 1.10 & 1.58 & 1.88 & 1.82 & 1.08 & 0 \\
\hline

\label{tab:tab1}
\end{tabular}
\end{table}

\begin{table}[h]
\centering
\caption{2$\alpha$ at different twist-angles in different graphene/TMDC heterostructures}
\begin{tabular}{cccccccc}
\hline
 2$\alpha(\mathrm{meV})$& 0\textdegree & 5\textdegree & 10\textdegree & 15\textdegree & 20\textdegree & 25\textdegree & 30\textdegree \\
\hline
$\mathrm{MoS_2}$  & 0.21 & 0.45 & 0.68 & 0.70 & 0.41 & 0.17 & 0.06 \\
\hline
$\mathrm{WS_2}$  & 0.32 & 0.80 & 1.37 & 1.54 & 0.78 & 0.31 & 0.11 \\
\hline
$\mathrm{MoSe_2}$  & 0.54 & 0.62 & 0.78 & 1.02 & 1.44 & 1.65 & 1.51 \\
\hline
$\mathrm{WSe_2}$  & 0.77 & 0.98 & 1.31 & 1.62 & 1.76 & 1.56 & 1.31 \\
\hline

\label{tab:tab2}
\end{tabular}
\end{table}



\bibliographystyle{unsrtnat} 
\bibliography{final}


%\begin{thebibliography}{00}



%\end{thebibliography}
\end{document}